\documentclass[a4paper,conference]{IEEEtran}
\IEEEoverridecommandlockouts
\usepackage{cite}

\usepackage{amsmath,amssymb,amsfonts}
\usepackage{algorithmic}
\usepackage{graphicx}

\usepackage[absolute]{textpos}

\usepackage{textcomp}
\usepackage{url}
\usepackage{enumitem}
\setlist[itemize]{leftmargin=*}
\usepackage{booktabs}
\usepackage{multirow}
\usepackage{tabularx}

\usepackage{tikz}
\usetikzlibrary{positioning}

\usepackage{xcolor}
\def\BibTeX{{\rm B\kern-.05em{\sc i\kern-.025em b}\kern-.08em
    T\kern-.1667em\lower.7ex\hbox{E}\kern-.125emX}}

\usepackage{units}

\newcommand{\anonymous}[2]{}
\renewcommand{\anonymous}[2]{{{#2}}}

\newcommand{\copyrightstatement}{
    \begin{textblock}{15}(0.5,0.1)    
         \noindent
         \centering
         \textblockcolour{white}
         \footnotesize
         \copyright 2022 IEEE. Personal use of this material is permitted. Permission from IEEE must be obtained for all other uses, in any current or future media, including reprinting/republishing this material for advertising or promotional purposes, creating new collective works, for resale or redistribution to servers or lists, or reuse of any copyrighted component of this work in other works
    \end{textblock}}

\begin{document}

\title{Modeling of Energy Consumption and Streaming Video QoE using a Crowdsourcing Dataset}


\copyrightstatement

\author{
\IEEEauthorblockN{Christian Herglotz\IEEEauthorrefmark{1}, Werner Robitza\IEEEauthorrefmark{2}\IEEEauthorrefmark{3}, Matthias Kränzler\IEEEauthorrefmark{1}, Andre Kaup\IEEEauthorrefmark{1}, and Alexander Raake\IEEEauthorrefmark{3}}
\IEEEauthorblockA{\IEEEauthorrefmark{1}Friedrich-Alexander-Universität, Erlangen, Germany --
Email: \emph{firstname.lastname}@fau.de}
\IEEEauthorblockA{\IEEEauthorrefmark{2}AVEQ GmbH, Vienna, Austria --
Email: werner.robitza@aveq.info}
\IEEEauthorblockA{\IEEEauthorrefmark{3}Audiovisual Technology Group, TU Ilmenau, Germany --
Email: \emph{firstname.lastname}@tu-ilmenau.de}
}

\IEEEoverridecommandlockouts
\IEEEpubid{\makebox[\columnwidth]{978-1-6654-8794-8/22/\$31.00
\copyright 2022 IEEE \hfill} \hspace{\columnsep}\makebox[\columnwidth]{ }}

\maketitle

\begin{abstract}
In the past decade, we have witnessed an enormous growth in the demand for online video services. Recent studies estimate that nowadays, more than 1\% of the global greenhouse gas emissions can be attributed to the production and use of devices performing online video tasks. As such, research on the true power consumption of devices and their energy efficiency during video streaming is highly important for a sustainable use of this technology. At the same time, over-the-top providers strive to offer high-quality streaming experiences to satisfy user expectations. Here, energy consumption and QoE partly depend on the same system parameters. Hence, a joint view is needed for their evaluation. In this paper, we perform a first analysis of both end-user power efficiency and Quality of Experience of a video streaming service. We take a crowdsourced dataset comprising 447,000 streaming events from YouTube and estimate both the power consumption and perceived quality. The power consumption is modeled based on previous work which we extended towards predicting the power usage of different devices and codecs. The user-perceived QoE is estimated using a standardized model. Our results indicate that an intelligent choice of streaming parameters can optimize both the QoE and the power efficiency of the end user device. Further, the paper discusses limitations of the approach and identifies directions for future research.
\end{abstract}

\begin{IEEEkeywords}
Quality of Experience, video streaming, energy consumption, energy efficiency
\end{IEEEkeywords}

\begin{tikzpicture}[overlay, remember picture]
\path (current page.north) node (anchor) {};
\node [below=of anchor] {2022 14th International Conference on Quality of Multimedia Experience (QoMEX)};
\end{tikzpicture}

\section{Introduction}
\label{sec:introduction}
Due to the availability of portable devices, high-bandwidth wireless Internet connections, and a plethora of online video services, video streaming has become an integral part of the daily life of billions of people. In a recent study it was found that as a consequence, $1\%$ of the global greenhouse gas emissions could be attributed to online video services in 2017 \cite{ShiftFull19}.

A challenge in determining the global overall energy consumption of online video services lies in using the correct underlying data. 
Even for a single service, many assumptions have to be made -- for instance about the type of device or the screen resolution. Hence, the use of a dataset with real-world data is crucial for representative results. Also, the specific impact of different technical factors (e.g., video codecs) on the required power has to be known to make accurate predictions. These parameters are also relevant for the Quality of Experience (QoE) of streaming. We hypothesize that there is a tradeoff between ensuring good quality and achieving power efficiency, and that optimizations can be made in both aspects. To understand these relationships, a joint analysis of the underlying factors is required.

In this paper, we perform a first combined analysis of power consumption and QoE for YouTube video streaming. We construct a model for the power used for streaming on desktop/laptop clients, incorporating factors such as video codec, resolution, and bitrate. We calculate power statistics on the actual streaming use of typical end users based on a large dataset from \anonymous{anonymous}{\cite{Robitza2020}}. We then relate the power consumption to the QoE as measured through a standardized QoE model. Finally, we calculate theoretical potential power savings while keeping the QoE constant. 

As there are many unknowns along a typical video streaming transmission path, we cannot focus on the entire end-to-end chain. In particular, details about how services encode their videos are not known, as are characteristics of the transmission networks. As a first step in calculating the overall energy consumption, this paper therefore covers the client (end-user) side only, where sufficient data is available. Other aspects will be part of future work.

This paper is structured as follows:
Section~\ref{sec:related-work} presents background work.
Section~\ref{sec:dataset} describes the used dataset in more detail.
In Section~\ref{sec:modeling}, we provide background on the power modeling and quality modeling per video stream.
Section~\ref{sec:results} presents the results of our modeling. 
In Section~\ref{sec:discussion} we discuss the results and their limitations, and Section~\ref{sec:conclusion} concludes the paper.

\section{Related Work}
\label{sec:related-work}

In the past years, there has been a quickly growing interest in the analysis, modeling, and improvement of the energy and power efficiency of video streaming solutions. A large body of research targeted the energy efficiency of end-user devices such as smartphones, laptops, or desktop PCs. For example, detailed analyses of the power consumption of smartphones were performed in \cite{Carroll13,Herglotz20}. Further work aimed at optimizing the power consumption of these devices \cite{Li12,  Kim15}, where some of the approaches were adopted in standardization activities \cite{GREEN-MPEG}. In addition, efforts were undertaken to model and optimize the energy consumption of the video decoder \cite{Mallikarachchi20,Herglotz18,Herglotz19}, which is an essential part of the processing pipeline in end-user devices. 

Concerning data centers used by video providers, research has concentrated on content delivery networks (CDNs), where energy analysis and modeling was proposed in \cite{Bianco16,Goudarzi20}. It was found that the overall energy depends on the number of servers, the amount of video data, the number of streaming requests, and the encoding of videos. The energy consumption of transmission networks was examined in detail in \cite{Malmodin20,Coroama13}, where it was found that the energy consumption is mainly determined by the number of network nodes and the amount of transmitted data. In general, it was found that the energy efficiency of data centers and transmission networks has increased with the development of new technology \cite{Malmodin20}. 

In terms of data sources to estimate energy consumption for a total set of users, commercial video analytics providers such as Conviva regularly publish reports summarizing video streaming usage and performance-related indicators (e.g., \cite{Conviva2022}). These are often presented in a global fashion, providing results from multiple (unknown) platforms. Usage of services is broken down by device or technology, and general overview statistics of viewing habits can be derived from such reports. However, obtaining more detailed data for a certain market or platform is more challenging. Researchers therefore turn to crowdsourcing~\cite{Hossfeld2014} to establish their own datasets. While in the context of streaming, crowdsourcing has primarily been used to obtain individual user's ratings within dedicated tasks, recently, the approach has been extended for automatic, large-scale data collection. As an example, \cite{Nam} presented a solution using a web browser extension that tracked users' YouTube viewing behavior and performance. A similar technology was shown in \cite{Robitza2020}, whose dataset serves as the basis for our research.

Approaches like \cite{bezerra2017qoe} have estimated energy consumption for video streaming on smartphones in relation to quality. However, as a proxy for QoE, the Peak-Signal-to-Noise (PSNR) metric has been used, which is known to correlate poorly with subjective QoE. PSNR also does not incorporate the detrimental effects of quality switches and stalling events on QoE. The authors of \cite{ballesteros2016energy} have also measured energy consumption of mobile devices and related it with achievable QoE for video streaming -- here measured via subjective experiments. The authors found potential to improve energy efficiency depending on whether frames are skipped during a freezing (stalling) period.

\section{Dataset}
\label{sec:dataset}

For an accurate modeling of energy consumption of a large audience, we need to be able to break down the energy consumption calculation to a per-session level. To obtain valid and realistic patterns of video streaming use, we therefore base our work on the dataset published in \cite{Robitza2020}. It consists of 477,000 desktop-based streaming sessions from YouTube users in Germany. The data was collected throughout the year 2019 via recruitment within a crowdsourcing campaign. Users installed a custom measurement software that monitored all YouTube playbacks in the browser. 

For each video playback, the following information is available:
device type (laptop, PC);
video and audio bitrates, codecs, resolution, framerate;
initial loading delay and stalling events;
operating system type and version;
browser type and version;
screen and browser resolution;
video metadata such as duration and viewing count;
video QoE (details given in next section).

The entire dataset consists of around 1,700 days worth of streaming from 3,600 unique installations of the software. The average playback lasts about 5.5\,min, with an interquartile range of 4.77\,min. About half of the playbacks originated from laptops, the other half from PCs. 68.8\% of users used Chrome, the remainder Firefox.


\section{Modeling Power and QoE}
\label{sec:modeling}


\subsection{Power Modeling}
\label{sec:power-modeling}

We base our power model on a study performed for smartphones~\cite{Herglotz20}. Since in this work, we focus on laptops and desktop PCs, we conducted a new set of dedicated measurements for those devices and adapted the model accordingly. In \cite{Herglotz20}, we examined the dependency of a smartphone's power consumption on a variety of variables. These variables are properties of the video streaming process and can be categorized as follows: hardware configurations (e.g., state of network, display brightness), software configuration (e.g., different software implementations, configuration of the application such as hardware or software decoding), streaming parameters (e.g., type of network, overall bitrate), and video parameters (e.g., video bitrate, frame rate, resolution, video codec). With these, a generic, linear power model of the form 
\begin{equation}
    \hat P = \boldsymbol{v}\cdot \boldsymbol{x}
    \label{eq:powerModel}
\end{equation}
was developed, where $\hat P$ is the estimated mean power of the device, $ \boldsymbol{v}$ is a row vector of variables representing the streaming process as explained above, and $\boldsymbol{x}$  is a column vector containing the set of parameters that describe the impact of a variable on the power consumption. 

For this study, we adopted the set of variables from \cite{Herglotz20}, dropped variables that are not relevant for laptops and desktop PCs, and performed corresponding measurements for one representative device each. The set of variables that we consider in this paper is listed in Table~\ref{tab:trainedParams} (see Section~\ref{sec:modeling}). 



The constant offset models the idle power a device consumes, even if no process is running. 
The video bitrate $b_\mathrm{v}$, the video frame rate $f_\mathrm{v}$, and the pixels per second $G_\mathrm{v}$ are variable parameters of the video stream and mainly determine the decoder's power consumption. The video bitrate $b_\mathrm{v}$ is the mean bitrate over the entire video sequence, where only the video stream is considered (no audio or side information).

The Internet connection offset $F_\mathrm{online}$ and the bitrate $b_\mathrm{online}$ model the power consumption of the Wi-Fi receiver for the laptop and the Ethernet connection for the PC. The Internet connection offset indicates for the laptop whether Wi-Fi is enabled and connected to a network, such that the corresponding power can be attributed to the idle power of the Wi-Fi module. For the desktop PC, we assume that the device is always connected to the Internet via Ethernet such that the corresponding power is part of the constant offset ($N/A$ in Table~\ref{tab:trainedParams}). The online bitrate $b_\mathrm{online}$ includes the video bitrate $b_\mathrm{v}$ mentioned above with additional audio bitrate and side information. This bitrate corresponds to the mean bitrate over the entire sequence, too. 

The codec is defined as a binary variable $C\in\{0,1\}$ indicating H.264/AVC or VP9, respectively. Note that in the crowdsourcing dataset, 87\% of all playbacks used VP9, and 13\% H.264.\footnote{The main discriminatory variables influencing the choice of codecs were the number of views and likes per video, suggesting that YouTube had rolled out VP9 to more popular videos first to achieve the most gains from the reduced traffic that a more efficient codec offers, considering the higher CPU impact of encoding with VP9 compared to H.264.} In the following, these variables are summarized in the variable vector $\boldsymbol{v}$. 

Note that further parameters could be considered. For example, the display brightness or the state of other modules (Bluetooth, camera, low-power operating mode for battery saving, etc.) can change the overall power consumption. However, the state of these parameters is unknown within the crowdsourcing dataset such that default values are used in this study. In addition, it is worth mentioning that in the crowdsourcing dataset, high correlations between variables may be present, for example for the video bitrate and Internet bitrate. To simplify measurements for the training process, we split up these parameters, because in this way, we can perform most measurements with local videos instead of Wi-Fi-streamed videos. The advantage is that measuring local videos is faster because of more stable and reliable power measurements that are not disturbed by potential Wi-Fi fluctuations. 

\subsubsection{Power Measurement Setup}

To train the parameters $\boldsymbol{x}$, we constructed a dedicated power measurement setup, which was used for both the laptop and the desktop PC. As we are interested in the overall power consumption of the devices, we measured the power through the main power supply ($220\,$V AC). To make sure that the power consumption of the laptop was measured correctly, we ensured that the battery was fully charged during the measurement such that the power supply was entirely used for processing.  For the desktop PC, we performed an additional measurement for the display, which is usually powered by a separate supply. The laptop is a Microsoft Surface Book 2 (Intel Core i5-8350U CPU, $13.5$ inch LCD display at a resolution of $3000\times 2000$ pixels), the desktop PC uses an Intel Core i5-4690 CPU @ $3.5\,$GHz, and the display is a Fujitsu P24-8 WS Neo with a resolution of $1920\times 1200$ pixels at a frame rate of $60\,$Hz. 


To obtain mean values for the streaming power, we prepared a test set of videos by encoding 13 sequences from the JVET common test conditions (CTC) \cite{JVET_N1010} using FFmpeg at medium preset with four constant rate factors (18, 23, 28, 33) for H.264 and four target bitrates ($500\,$kbps, $1\,$Mbps, $2\,$Mbps, $4\,$Mbps) for VP9. Resolutions varied from $416\times240$ to $1920\times1080$, and frame rates from 24--60\,Hz.
While stalling is part of the crowd-based dataset, for this preliminary study we have ignored stalling in our power model because it affects streaming duration and would falsify measurements of instantaneous power. We will consider it in future work using, e.g., the overall energy consumption of a streaming session, where the streaming duration consists of both video playback and stalling.

The resulting bit streams, which have a fixed duration of $10\,$s, were multiplexed in MP4 
or WebM containers, where in each container, the sequence was repeated four times. 
Hence, we could then measure the mean power during the playback of the $10\,$s-sequence, although the exact beginning and end of the playback process was unknown. We set the power meter integration time to $10\,$s and triggered the measurement roughly $10\,$s after starting the playback. This method ensured that energy which is needed to launch the video playback application did not interfere with our measurement. On the laptop, we used the VLC player 
with Direct3D$\,$11 
hardware acceleration for H.264-coded videos and software decoding for VP9-coded videos, such that in general, a higher power consumption for VP9-coded videos can be expected \cite{Khernache21}. On the PC, we used \texttt{ffplay} with software decoding for both codecs. 

The measurements indicate that during video playback (online and offline), the lower and the upper bound of measured power values ranges between $10\,$W and $20\,$W for the laptop and between $80\,$W and $110\,$W for the PC, including the display. 

\subsubsection{Energy Model Training}
In order to determine model parameters, we collect all measured powers for both devices in the column vectors $\boldsymbol{P}_\mathrm{PC}$ and $\boldsymbol{P}_\mathrm{laptop}$ of size $M$, which corresponds to the number of measurements. In the following, we use the index $d\in\{\mathrm{PC},\mathrm{laptop}\}$ to indicate both kinds of devices. Furthermore, we construct the variable matrix $\boldsymbol{V}$ of size $M\times L$, where $L$ is the number of power modeling variables from Table~\ref{tab:trainedParams}. Note that $\boldsymbol{v}$ from \eqref{eq:powerModel} corresponds to a single row of $\boldsymbol{V}$.  The least-squares optimal parameter values $\boldsymbol{x}_{d}$ of size $L$ are then determined by calculating the Moore-Penrose inverse of $\boldsymbol{V}$ and rewriting (\ref{eq:powerModel}) to 
\begin{equation}
    \boldsymbol{x}_d = \boldsymbol{P}_d\cdot \boldsymbol{V}^{-1}. 
    \label{eq:invPowerModel}
\end{equation}
The mean estimation errors, which are calculated by 
\begin{equation}
    \bar e_d = \frac{1}{M} \sum_{m=1}^M \frac{\left|\boldsymbol{v}_m\cdot \boldsymbol{x}_d - {P}_{d,m}\right|}{{P}_{d,m}}, 
\end{equation}
where $\boldsymbol{v}_m$ is the $m$-th row of $\boldsymbol{V}$ and $P_{d,m}$ is the $m$-th entry of $\boldsymbol{P}_d$, 
were found to yield $\bar e_\mathrm{PC}= 1.75\%$ and $\bar e_\mathrm{laptop}= 1.12\%$, respectively, which is even lower than values reported for smartphones in \cite{Herglotz20}. Note that these mean estimation errors were determined using 10-fold cross-validation. 

\begin{figure}[tb]
    \centering
    \includegraphics[width=1\linewidth]{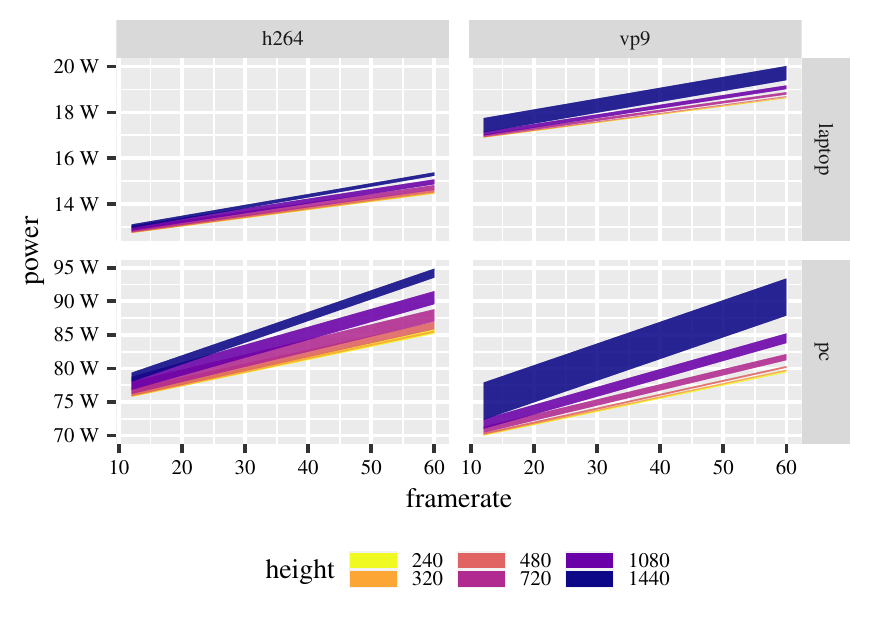}
    \vspace{-0.8cm}
    \caption{Impact of frame rate (horizontal axis), video resolution (line color), and bitrate (width of bands)  on power consumption (vertical axis).}
    \label{fig:model-characteristic}
    \vspace{-0.5cm}
\end{figure}

The training returned the parameter values shown in Table~\ref{tab:trainedParams}. Next to the trained values, we indicate the maximum impact of a certain variable $V$ on device $d$ given by 
\begin{equation}
I_{d,V} = x_{d,V} \cdot V_\mathrm{max}. 
\end{equation}
For example, the impact of the frame rate $I_{d,f_\mathrm{v}}$ is calculated by multiplying $x_{d,f_\mathrm{v}}$ with the maximum considered frame rate of $V_\mathrm{max}=f_\mathrm{v,max}=60\,$Hz. Figure~\ref{fig:model-characteristic} visualizes this impact graphically, split by device and codec, using the frame rate and resolution as two major parameters. The width of the bands corresponds to the possible variation due to different bitrates. The typical bitrate ranges were determined from the crowd dataset.

We can see that as expected, the values for the desktop PC are significantly higher than for the laptop, which reflects the range of values mentioned in the last subsection. For the laptop, next to the constant offset, the choice of the video codec has the highest impact on the power consumption. On the PC, the frame rate and the bitrate show the highest impact. 

Interestingly, using VP9, the laptop consumes more than $4\,$W of additional power. In contrast, the PC consumes $5.73\,$W less on average when using VP9. The reason for these differences depending on the codec can be attributed to the choice of the decoder implementation (hardware accelerated, software, and code optimizations). Hence, these results must be understood as first indications of expected power ranges only. Still, they show that the choice of the codec can have a significant impact on the power consumption.

Furthermore, the negative value for the online streaming bitrate on the PC ($x_{b_\mathrm{online}}$) seems counter-intuitive. In this case, this negative impact must be considered jointly with the offline bitrate parameter $x_{b_\mathrm{v}}$, which has a positive value that is greater than the absolute value of $x_{b_\mathrm{online}}$. An explanation for this behavior can be that downloading through the Ethernet connection consumes less power than loading from the local hard drive.

\begin{table*}[t]
\caption{Trained parameters for the power model \eqref{eq:powerModel}.
pps = number of displayed full color pixels per second.}
\label{tab:trainedParams}
\vspace{-0.3cm}
\begin{center}
\begin{tabular}{r @{\hskip 0.9cm} l @{\hskip 0.9cm} r @{\hskip 0.9cm} r @{\hskip 0.9cm} r @{\hskip 0.9cm} r}
\toprule
 & & \multicolumn{2}{c}{$d=\,$\textbf{Laptop}} & \multicolumn{2}{c}{$d=\,$\textbf{Desktop PC}}\\
\textbf{Variable $V$} & \textbf{Meaning} & Trained Values $\boldsymbol{x}_d $ & Max. Impact $I_{d,V}$ & Trained Values $\boldsymbol{x}_d$ & Max. Impact $I_{d,V}$ \\
\midrule
$1$ & Constant offset  & $8.52\,$ W& $8.52\,$ W& $73.3\,$ W& $73.3\,$ W\\

$f_\mathrm{v}$ & Video frame rate& $0.035\,\nicefrac{\mathrm{W}}{\mathrm{fps}}$ & $2.12\,$ W 
& $0.19\,\nicefrac{\mathrm{W}}{\mathrm{fps}}$& $11.5\,$ W\\

$G_\mathrm{v}$ & Pixels per second& $5.76\cdot 10^{-9}\,\nicefrac{\mathrm{W}}{\mathrm{pps}}$& $0.72\,$ W
& $6.29\cdot 10^{-8}\,\nicefrac{\mathrm{W}}{\mathrm{pps}}$& $7.82\,$ W\\

$b_\mathrm{v}$ & Video bitrate & $6.97\cdot 10^{-6}\,\nicefrac{\mathrm{W}}{\mathrm{kbps}}$& $0.13\,$ W
&$4.05\cdot 10^{-4}\,\nicefrac{\mathrm{W}}{\mathrm{kbps}}$ & $7.29\,$ W\\

$b_\mathrm{online}$ &{Bitrate (Internet)}& $3.24\cdot 10^{-5}\,\nicefrac{\mathrm{W}}{\mathrm{kbps}}$ & $0.58\,$ W
& $-5.28\cdot 10^{-5}\,\nicefrac{\mathrm{W}}{\mathrm{kbps}}$& $-0.95\,$ W\\

$F_\mathrm{online}$& Internet conn. offset & $1.15\,$W & $1.15\,$ W 
& \emph{N/A} & \emph{N/A}\\


$C$ & {Video codec}& $4.16\,$W& $4.16\,$ W
& $-5.73\,$ W& $-5.73\,$ W\\
\bottomrule
\end{tabular}
\end{center}
\vspace{-0.5cm}
\end{table*}



\subsection{Quality Model}

To get an estimate of the quality of every single video playback, we used the standardized ITU-T Rec. P.1203 QoE model. It has been trained and validated for the HTTP Adaptive Streaming use case and considers the impact of initial loading delay, stalling, and quality variation throughout the playback. The output of the model is a Mean Opinion Score (MOS) from 1 to 5 (1 = bad, 5 = excellent). We calculated the MOS results with an open-source reference software \cite{Raake2017,Robitza2018}.\footnote{\url{https://github.com/itu-p1203/itu-p1203}} Since the standardized model does not have support for VP9, we used a publicly available extension to obtain the results for the respective playbacks.\footnote{\url{https://github.com/Telecommunication-Telemedia-Assessment/itu-p1203-codecextension}} Also, when used strictly according to standard, the model is only valid for playbacks of at most 5\,min length. Since this matched the average video session duration in our dataset, we calculated the MOS score for the first 5\,min of every sequence only.

\section{Results}
\label{sec:results}



\begin{figure}[tb]
    \centering
    \includegraphics[width=1\linewidth]{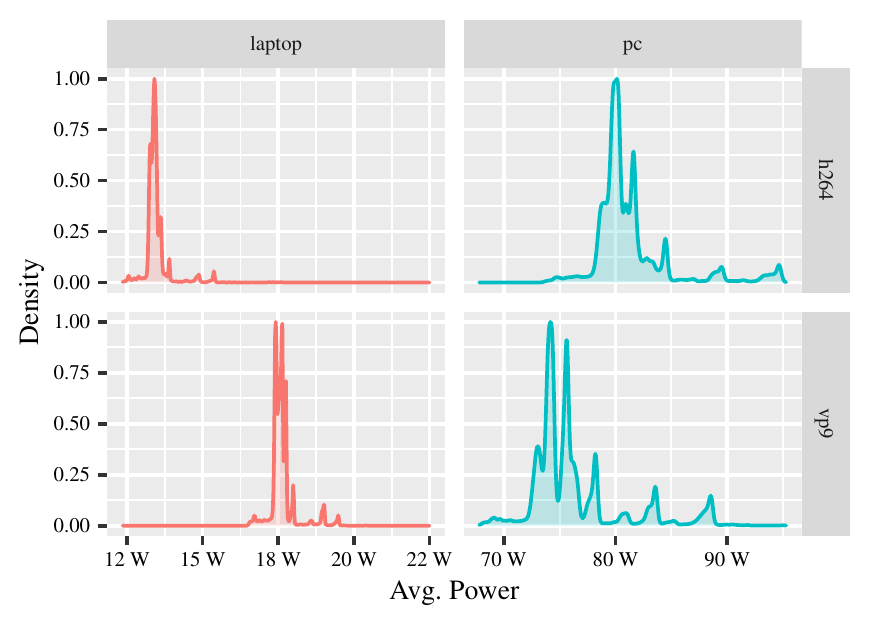}
    \vspace{-0.8cm}
    \caption{Avg. power used for PCs/laptops and different codecs.}
    \label{fig:power-avg}
    \vspace{-0.5cm}
\end{figure}

Looking at how much power was consumed within the dataset, we first estimated the average power for each streaming session using the model from Section~\ref{sec:power-modeling}. The sessions were truncated to the first five minutes, which corresponds to the validity of MOS values from P.1203. The results are shown in Figure~\ref{fig:power-avg}, grouped by device and codec. The average for laptops is $13.5\,$W and $17.6$\,W for H.264 and VP9, respectively. For PCs, it is $81.3\,$W and $76.3$\,W, respectively.

In terms of QoE, generally, playback quality is good: the MOS score is 4.12 on average, with an interquartile range (IQR) of 0.55. The peak is at 4.57; a second peak can be observed at 4.27, explained by the occurrence of one subsecond stalling event for 22\% of video playbacks, as reported by the YouTube API. 



\begin{figure}[tb]
    \centering
    \includegraphics[width=1\linewidth]{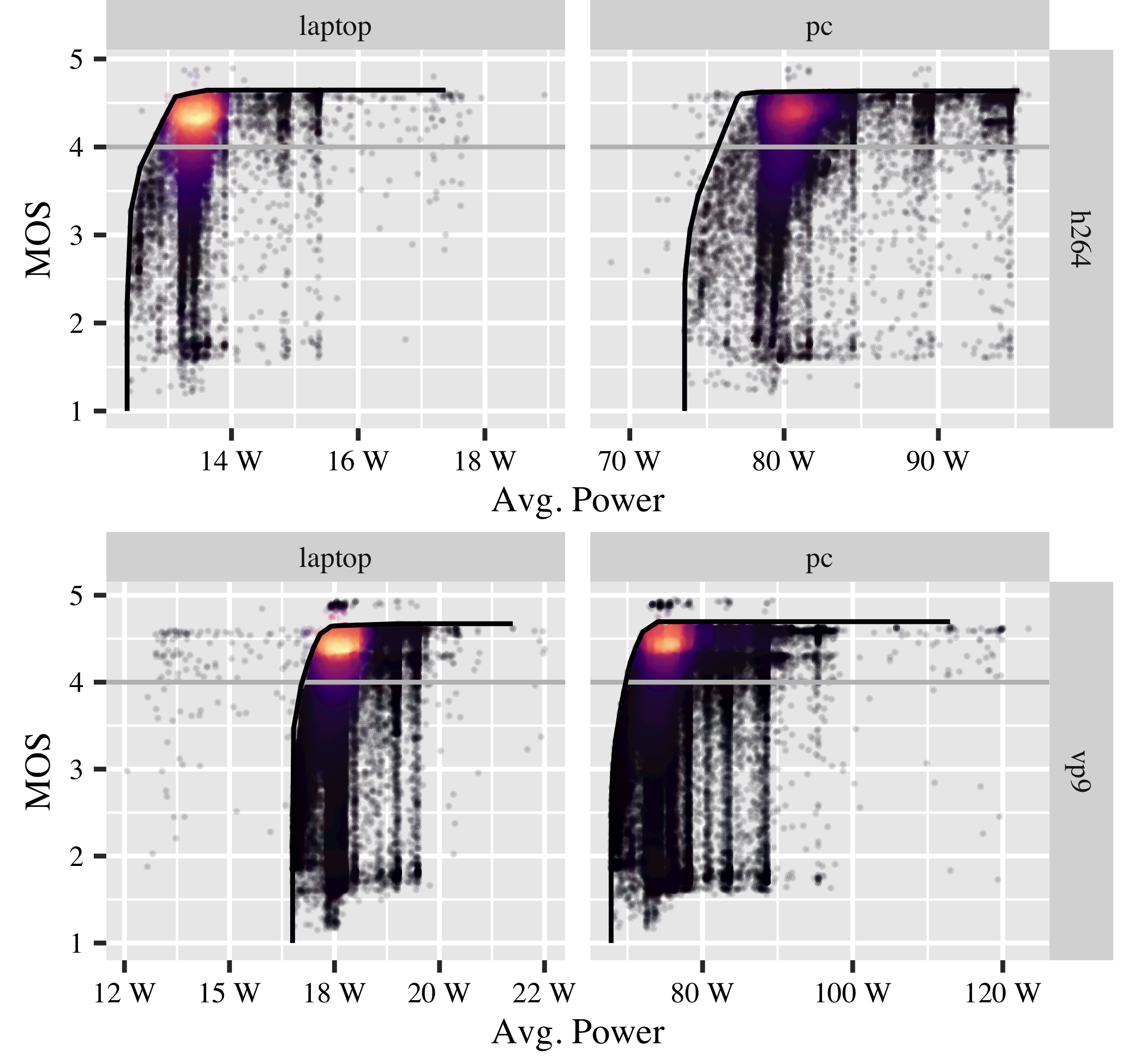}
    \vspace{-0.8cm}
    \caption{Avg. power vs MOS for PCs/laptops and different codecs. Brighter color implies higher density. The black line indicates the convex hull. }
    \label{fig:power-vs-mos}
    \vspace{-0.5cm}
\end{figure}



Figure~\ref{fig:power-vs-mos} shows the distribution of MOS values against the average used power. Lighter areas indicate a higher density of points. We have added two lines for guidance: first, the threshold of MOS = 4 indicates sessions that are good or better. This also corresponds to 75\% of all sessions. Second, a convex hull has been computed, showing us the optimal set of points in terms of achieving the best quality at the lowest possible power. For its calculation, outliers have been removed.


The distribution of MOS values and power values of the streaming sessions in Fig.~\ref{fig:power-vs-mos} shows that the highest density of points can be observed at the top left of the plots, close to a ``sweet spot'' of the convex hull representing rather low power and high MOS values (e.g., below $14\,$W and above a MOS of $4$ for H.264 decoding on the laptop, respectively). This shows that in general, video parameters are often chosen in a power-efficient and high-quality manner.  However, the distribution also shows that many sessions are located far from the convex hull. 
The separation between vertical clusters of lines mainly corresponds to different video resolutions being streamed. Low MOS values in this case are due to initial loading or stalling (which is not yet reflected in the power model). Still, for sessions with a high MOS (and consequently no stalling), this indicates that at the same overall quality, a significant amount of power could be saved on the end-user side choosing optimized video parameters (i.e. those that optimize for the P.1203-estimated QoE). In the following, we discuss several methods to optimize the power consumption and calculate average potential savings. 

First, we investigate the potential power savings if all streaming sessions (except outliers) were located on the convex hull. For this, we linearly interpolate the convex hull as a function given by 
\begin{equation}
P^\mathrm{hull}_{d,c}(s) = f(\mathrm{MOS}_s),
\end{equation}
where $d$ is the device, $c$ is the codec, and $s$ the index of the current session. Then, we calculate the optimized power for each streaming session $s$ and compare it with the estimated power $P_s$ and obtain absolute and average savings as  
\begin{equation}
    \Delta P_{d,c,s} = P^\mathrm{hull}_{d,c}(s) - P_s ;  \quad \delta P_{d,c,s} = \frac{P^\mathrm{hull}_{d,c}(s) - P_s}{P_s},
\end{equation}
respectively. Averaging over all sessions as
\begin{equation}
    \Delta P_{d,c} = \frac{1}{\left|S_{d,c}\right|}\sum_{s\in S_{d,c}}\Delta P_{d,c,s};\quad \delta P_{d,c} = \frac{1}{\left|S_{d,c}\right|}\sum_{s\in S_{d,c}}\delta P_{d,c,s} ,
\end{equation}
we find mean power savings for each device $d$ and codec $c$ as listed in Table~\ref{tab:savings}, where $S_{d,c}$ is the set of sessions on device $d$ using codec $c$. 
\begin{table}[t]
\caption{Average power savings when choosing power-optimal parameters for each streaming session.}
\label{tab:savings}
\vspace{-0.3cm}
\begin{center}
\begin{tabular}{r @{\hskip 0.7cm} r @{\hskip 0.7cm} r @{\hskip 0.7cm} r @{\hskip 0.7cm} r}
\toprule
 Device $d$ & \multicolumn{2}{c}{\textbf{Laptop}} & \multicolumn{2}{c}{\textbf{Desktop PC}}\\
Codec $c$ & H.264 & VP9 & H.264 & VP9  \\
\midrule
$\Delta P_{d,c}$ & $-4.71\%$ & $-3.82\%$ & $-6.52\%$ & $-6.95\%$ \\
$\delta P_{d,c}$ & $-0.64\,$W & $-0.68\,$W & $-5.41\,$W & $-5.49\,$W\\
\bottomrule
\end{tabular}
\end{center}
\vspace{-0.5cm}
\end{table}

We can see that on both devices and for both codecs, significant average power savings of more than $3.5\%$ can be obtained. Maximum savings were found to reach more than $20\%$ for all devices. Note that these power savings can be obtained by changing the resolution, the frame rate, and the bitrate of the videos. 

Furthermore, we can see that the choice of the codec and the decoding technology can have a significant impact on the power consumption. For example, in our scenario, switching from VP9 to H.264 for all sessions on the laptop and optimizing using the convex hull, we obtain mean power savings of $26.8\%$ or $4.75\,$W. When switching from PC to laptop with H.264 decoding, we can even save $84\%$ or $68.5\,$W without loss in visual quality. In future work, we will exploit this knowledge in order to develop a method enabling convex-hull video streaming for energy efficiency on the client side. 


\section{Discussion and Limitations}
\label{sec:discussion}

This work could only cover a part of the entire end-to-end streaming chain. We based our power savings calculations on a first set of measurements conducted on a specific laptop and PC, since there exists no generic model for client-side streaming power usage. The results indicate that the availability of hardware decoding capabilities significantly impacts the resulting power demands, and that for a valid extrapolation to a larger set of users, such data cannot be directly used. Hence, it will be necessary to extend the laboratory measurements to a broader set of devices, using browser-based measurements (e.g., with dash.js) instead of standalone software. Also, one will need to gather more information about typical end-users' device capabilities, where a wide variety of browsers (including legacy versions), operating systems, codecs, and CPU determine the final power usage. Here, another crowdsourcing study may help in quantifying typical device capabilities. That said, additional device- and context-specific features like low-energy modes in smartphones may complicate the goal of finding a universal model for energy consumption.


In terms of QoE, the used dataset and quality model provided a good first basis for calculating possible power savings, since the measurements contained realistic and representative variations of playback quality. A possible limitation could be the constant change in how OTTs provide their streams, leading to new codecs that have to be considered (e.g., AV1 in the case of YouTube), and different underlying video encoding parameters.

\section{Conclusion}
\label{sec:conclusion}
In this paper, we introduced a novel method for a joint analysis of power consumption and QoE. We showed that for OTT video providers, there is a huge potential in optimizing the overall energy consumption of end-user devices while keeping QoE high. In this respect, parameters to optimize for include the video codec (e.g., choosing a more power-efficient codec), resolution, frame rate, and bitrate. In our measurements, possible power savings for a given device yielded at least $3.5\%$.

However, finding optimal parameters highly depends on the user device and must be optimized in a per-session manner. One must also keep in mind that a) high QoE and power savings are partly conflicting goals, and b) the end-to-end view must be considered. For instance, the amount of overall traffic will be reduced with VP9 compared to H.264, even if its encoding and decoding may be less efficient for some devices. Knowledge about the device processing capabilities must therefore be considered from an end-to-end perspective. Also, implementing a system for optimizing energy consumption may introduce additional processing on top of regular service operation if not done efficiently.

In future work, we will investigate further devices, codecs, and decoder implementations, to create a more generic client-based power consumption model. A joint analysis with playback QoE will be performed on a per-session basis, to identify the optimal tradeoffs between saving energy and ensuring high QoE. 
Furthermore, we will extend our work to also consider the power consumption of OTT providers and ISPs. 

\bibliographystyle{IEEEbib}

\begin{thebibliography}{10}

\bibitem{ShiftFull19}
{The Shift Project},
\newblock ``Climate crisis: The unsustainable use of online video,''
\newblock Tech. {R}ep., 2019.

\bibitem{Robitza2020}
W.~Robitza, A.~Dethof, S.~G{\"{o}}ring, A.~Raake, T.~Polzehl, and A.~Beyer,
\newblock ``{Are You Still Watching? Streaming Video Quality and Engagement
  Assessment in the Crowd},''
\newblock in {\em 12th International Conference on Quality of Multimedia
  Experience (QoMEX 2020)}, Athlone, Ireland, 2020.

\bibitem{Carroll13}
A.~Carroll and G.~Heiser,
\newblock ``The systems hacker's guide to the galaxy - energy usage in a modern
  smartphone,''
\newblock in {\em Proc. 4th Asia-Pacific Workshop on Systems (APSys)},
  Singapore, 2013.

\bibitem{Herglotz20}
C.~{Herglotz}, S.~{Coulombe}, C.~{Vazquez}, A.~{Vakili}, A.~{Kaup}, and
  J.~{Grenier},
\newblock ``Power modeling for video streaming applications on mobile
  devices,''
\newblock {\em IEEE Access}, vol. 8, pp. 70234--70244, 2020.

\bibitem{Li12}
X.~Li, Z.~Ma, and F.~C.~A. Fernandes,
\newblock ``Modeling power consumption for video decoding on mobile platform
  and its application to power-rate constrained streaming,''
\newblock in {\em Proc. Visual Communications and Image Processing (VCIP)}, San
  Diego, USA, Nov. 2012.

\bibitem{Kim15}
J.-W. Kim, K.-H. Lee, J.-G. Bae, H.-G. Kim, and J.-O. Kim,
\newblock ``Optimal liquid crystal display backlight dimming based on clustered
  contrast loss,''
\newblock {\em Optical Engineering}, vol. 54, no. 10, pp. 103--112, 2015.

\bibitem{GREEN-MPEG}
{\em Green MPEG},
\newblock ITU-T Rec. H.265 and ISO/IEC 23008-2, ITU-T and ISO/IEC JTC 1/SC29/WG
  11 (MPEG), Apr 2013.

\bibitem{Mallikarachchi20}
T.~Mallikarachchi, D.~Talagala, H.~Kodikara~Arachchi, C.~Hewage, and
  A.~Fernando,
\newblock ``A decoding-complexity and rate-controlled video-coding algorithm
  for {HEVC},''
\newblock {\em Future Internet}, vol. 12, no. 7, pp. 120, 2020.

\bibitem{Herglotz18}
C.~Herglotz, D.~Springer, M.~Reichenbach, B.~Stabernack, and A.~Kaup,
\newblock ``Modeling the energy consumption of the {HEVC} decoding process,''
\newblock {\em IEEE Transactions on Circuits and Systems for Video Technology},
  vol. 28, no. 1, pp. 217--229, Jan. 2018.

\bibitem{Herglotz19}
C.~Herglotz, A.~Heindel, and A.~Kaup,
\newblock ``Decoding-energy-rate-distortion optimization for video coding,''
\newblock {\em IEEE Transactions on Circuits and Systems for Video Technology},
  vol. 29, no. 1, pp. 172--181, Jan. 2019.

\bibitem{Bianco16}
A.~{Bianco}, R.~{Mashayekhi}, and M.~{Meo},
\newblock ``Energy consumption for data distribution in content delivery
  networks,''
\newblock in {\em Proc. IEEE International Conference on Communications (ICC)},
  2016, pp. 1--6.

\bibitem{Goudarzi20}
P.~{Goudarzi}, A.~{Ghassemi}, M.~R. {Mirsarraf}, and R.~{Buyya},
\newblock ``Joint energy-{QoE} efficient content delivery networks using
  real-time energy management,''
\newblock {\em IEEE Systems Journal}, vol. 14, no. 1, pp. 927--938, 2020.

\bibitem{Malmodin20}
J.~Malmodin,
\newblock ``The power consumption of mobile and fixed network data services -
  the case of streaming video and downloading large files,''
\newblock in {\em Proc. Electronic Goes Green (EGG)}, 2020.

\bibitem{Coroama13}
V.~C Coroama, L.~M Hilty, E.~Heiri, and F.~M. Horn,
\newblock ``The direct energy demand of {I}nternet data flows,''
\newblock {\em Journal of Industrial Ecology}, vol. 17, no. 5, pp. 680--688,
  2013.

\bibitem{Conviva2022}
Conviva,
\newblock ``{Conviva's State of Streaming Q4/2021},''
\newblock Tech. {R}ep., 2022.

\bibitem{Hossfeld2014}
T.~Hossfeld, C.~Keimel, M.~Hirth, B.~Gardlo, J.~Habigt, K.~Diepold, and
  P.~Tran-Gia,
\newblock ``{Best practices for QoE crowdtesting: QoE assessment with
  crowdsourcing},''
\newblock {\em IEEE Transactions on Multimedia}, vol. 16, no. 2, pp. 541--558,
  2014.

\bibitem{Nam}
H.~Nam, K.-H. Kim, and H.~Schulzrinne,
\newblock ``{QoE Matters More Than QoS: Why People Stop Watching Cat Videos},''
\newblock in {\em Proc. IEEE International Conference on Computer
  Communications}, 2016.

\bibitem{bezerra2017qoe}
C.~Bezerra, A.~De~Carvalho, D.~Borges, N.~Barbosa, J.~Pontes, and E.~Tavares,
\newblock ``Qoe and energy consumption evaluation of adaptive video streaming
  on mobile device,''
\newblock in {\em 2017 14th IEEE Annual Consumer Communications \& Networking
  Conference (CCNC)}, 2017, pp. 1--6.

\bibitem{ballesteros2016energy}
L.~G.~M. Ballesteros, S.~Ickin, M.~Fiedler, J.~Markendahl, K.~Tollmar, and
  K.~Wac,
\newblock ``Energy saving approaches for video streaming on smartphone based on
  qoe modeling,''
\newblock in {\em 2016 13th IEEE Annual Consumer Communications \& Networking
  Conference (CCNC)}, 2016, pp. 103--106.

\bibitem{JVET_N1010}
F.~Bossen, J.~Boyce, X.~Li, V.~Seregin, and K.~S\"uhring,
\newblock ``{JVET} common test conditions and software reference configurations
  for {SDR} video,''
\newblock {AHG Report, JVET-N1010}, ITU/ISO/IEC Joint Video Exploration Team
  (JVET), {Jan.} 2017.

\bibitem{Khernache21}
M.~B.~A. Khernache, Y.~Benmoussa, J.~Boukhobza, and D.~Menard,
\newblock ``{HEVC} hardware vs software decoding: An objective energy
  consumption analysis and comparison,''
\newblock {\em Journal of Systems Architecture}, vol. 115, pp. 102004, 2021.

\bibitem{Raake2017}
A.~Raake, M.-N. Garcia, W.~Robitza, P.~List, S.~Göring, and B.~Feiten,
\newblock ``{A bitstream-based, scalable video-quality model for HTTP adaptive
  streaming: ITU-T P.1203.1},''
\newblock in {\em Proc. 9th International Conference on Quality of Multimedia
  Experience (QoMEX)}, Erfurt, May 2017, IEEE.

\bibitem{Robitza2018}
W.~Robitza, S.~Göring, A.~Raake, D.~Lindegren, G.~Heikkilä, J.~Gustafsson,
  P.~List, B.~Feiten, U.~Wüstenhagen, M.-N. Garcia, K.~Yamagishi, and
  S.~Broom,
\newblock ``{HTTP Adaptive Streaming QoE Estimation with ITU-T Rec. P.1203 –
  Open Databases and Software},''
\newblock in {\em Proc. 9th ACM Multimedia Systems Conference}, Amsterdam,
  2018.

\end{thebibliography}

\end{document}